\documentstyle[aps,prd,twocolumn,graphicx,floats]{revtex}

\begin{document}

\wideabs{
\begin{flushright}
LBNL-50255 \\
{\tt astro-ph/0205520} \\
\end{flushright}

\title{Domain Walls as Dark Energy}
\author{Alexander Friedland$^1$, Hitoshi Murayama$^{2,3}$, and Maxim Perelstein$^3$}
\address{
$^1$ School of Natural Sciences,
Institute for Advanced Study,
Einstein Drive,
Princeton, NJ 08540\\
$^2$Department of Physics, 
University of California, 
Berkeley, CA~~94720\\ 
$^3$Theory Group, 
Lawrence Berkeley National Laboratory, 
Berkeley, CA~~94720}

\date{December 4, 2002}
\maketitle

\begin{abstract}
  The possibility that the energy density of the Universe is dominated
  by a network of low-tension domain walls provides an alternative to
  the commonly discussed cosmological constant and scalar-field
  quintessence models of dark energy. We quantify the lower bound on
  the number density of the domain walls that follows from the
  observed near-isotropy of the cosmic microwave background radiation.
  This bound can be satisfied by a strongly frustrated domain wall network.
  No fine-tuning of the parameters of the underlying field theory model
  is required. We briefly outline the observational consequences of this
  model.  
\end{abstract}
}

\newpage

\section{Introduction}

In the last few years, there has been a growing amount of evidence
that an unknown negative pressure component (``dark energy'') accounts
for $70\pm 10$\% of the energy density of the Universe.  
Arguably the most direct evidence comes from the luminosity-redshift
relation of the Type Ia supernovae \cite{hzs,scp}.  Additional
evidence is provided by a combination of data \cite{concord}.  Both the
analysis of the formation of the large scale structure \cite{LSS} and
the studies of the baryon fraction in galaxy clusters \cite{clusters},
combined with the Big Bang Nucleosynthesis calculations \cite{BBN},
suggest the matter content significantly below the critical density.
On the other hand, the cosmic microwave background radiation (CMBR)
power spectrum suggests a flat Universe \cite{CMBfirstpeak}. Therefore
the difference must be made up by an additional non-clustering energy
density component.
The relationship between the pressure $p$ and the energy density
$\rho$ of the dark energy component is usually parameterized by
$p=w\rho$, where the equation of state $w$ could be time dependent.
Theoretical considerations prefer $w \ge -1$. A good fit to the
supernova data can be obtained assuming a constant or slowly changing
equation of state satisfying $w \lesssim -0.5$~\cite{concord} or $w
\lesssim -0.6$~\cite{PTW}, depending on the details of the analysis.

The precise nature of the dark energy component has been a subject of
intense theoretical speculation.  The most popular dark energy
candidates include the cosmological constant, or vacuum energy, and
the so-called quintessence, a nearly spatially homogeneous but
time-dependent scalar field (see, for example, 
Refs.~\cite{quintessence,tracker1,tracker2}.) Another very interesting 
possibility is the ``solid'' dark energy, originally suggested by
Bucher and Spergel \cite{BucherSpergel,Spergel}. In this case, one
postulates that the dark energy component possesses resistance to pure
shear deformations, guaranteeing stability with respect to small
perturbations. (The stability condition is nontrivial for substances
with negative pressure. It is violated, for example, by a perfect
fluid.) 

A possible microphysical origin for solid dark energy is a dense
network of low-tension domain walls~\cite{Spergel,Hill}.  This is
attractive for several reasons.  First, domain walls are ubiquitous in
field theory, inevitably appearing in models with spontaneously broken
discrete symmetries~\cite{Vilenkin}. Second, domain walls, and the
solid dark energy in general, have been shown to be compatible with
the observations of large scale structure~\cite{Spergel,Fabris}.
Finally, a static wall network has an equation of state $w=-2/3$
\cite{KT}, consistent with all observational data. Despite these
appealing features, the idea that a domain wall network could play the
role of dark energy has not received much attention in the literature.
In this paper, we investigate certain aspects of this scenario. We
show that the requirement of the isotropy of the CMBR yields a strong
lower bound on the number density of the domain walls. We then
consider the implications of this bound for the field-theoretic models
of the domain walls. We argue that the domain wall network has to be
strongly frustrated to satisfy the CMBR constraint. If this is the
case, the equation of state of dark energy is expected to be close to
$-2/3$.

\section{CMBR constraints}

It is well known that the Universe is spatially homogeneous and
isotropic on large scales. The best evidence for this is provided by
the observed near-isotropy of the CMBR. If a network of domain walls is
present, the CMBR photons would acquire different gravitational
redshifts (or blueshifts) depending on the direction of their
arrival~\cite{ZKO}. Let us estimate this effect.
 
Until $z\sim O(1)$, the contribution of the walls to the energy
density was subdominant and hence they did not significantly affect
the CMBR anisotropy. Most of the anisotropy due to domain walls would
be built up during the relatively recent epoch, $z \lesssim 1$. This
allows us to simplify the problem. Consider an observer at rest in the
comoving frame. (After subtracting the effects due to the peculiar
motion of the Earth, the CMBR anisotropy measured in actual
experiments corresponds to what would be measured by such an
observer.) Let $R$ denote the comoving distance corresponding to
$z=z_0<1$, and consider the sphere ${\cal S}$ of radius $R$ centered on 
the observer. We assume that the CMBR photons arriving at the surface of 
this sphere are isotropic. The anisotropy induced when the photons are 
travelling from the surface of ${\cal S}$ to the observer can be simply 
estimated, up to corrections suppressed by powers of $z_0$, using Newtonian 
theory. Indeed, the results of the general theory of relativity on 
subhorizon scales with receding velocities $v\ll c$ must be equally well 
described using Newtonian gravity. Recall that in the Newtonian picture, 
the observer is taken to be at ``the center of the Universe''. Homogeneous 
matter around him creates a gravitational field which grows with distance as
\begin{equation}
{\bf g}(r)=-G\frac{4\pi}{3}(\rho+3 p)r{\bf\hat{r}}.
\label{eq:homofield}
\end{equation}
Here we are following the general principle that the source of Newtonian 
gravity is the combination $(\rho+3 p)$ and not the density itself (see, for 
example, \cite{Peebles}). The redshifts of distant objects in this picture 
result from the Doppler shift due to the expansion of the Universe, and the 
effect of the gravitational field in Eq.~(\ref{eq:homofield}). In Appendix A, 
we discuss the calculation of these effects in a homogeneous Universe, and 
show that the Newtonian calculation is accurate up to and including the 
second order terms in the relative velocity of the emitter and the 
observer. 

If domain walls 
are present, an additional, anisotropic gravitational field will be induced 
on top of (\ref{eq:homofield}), leading to anisotropies in the CMBR  
temperature. Denoting the Newtonian potential of the wall network by
$\Phi ({\bf r})$, we can write the observed temperature difference
between points ${\bf 1}$ and ${\bf 2}$ on the surface of the sphere 
${\cal S}$ as
\begin{equation}
  \label{eq:deltaT}
  {T({\bf \hat{r}_1})-T({\bf \hat{r}_2}) \over \langle T \rangle}
= \Phi ({\bf r_1}) - \Phi({\bf r_2}),
\end{equation}
where ${\bf r_{1,2}}$ are the vectors pointing from the center of the
sphere to points ${\bf 1}$ and ${\bf 2}$ and ${\bf \hat{r}_{1,2}} = 
{\bf r_{1,2}}/R$.
 
 The Newtonian potential at point ${\bf x}$ from a single planar wall
 is given by \cite{Vilenkin1981}
\begin{equation}
  \label{eq:onewall}
  \phi({\bf x};\,{\bf\hat{n}}, {\bf x_0})= -2\pi G_N\sigma\;
{\bf\hat{n}} \cdot({\bf x}-{\bf x_0}),
\end{equation}
where $\sigma$ is the wall tension,  ${\bf x_0}$ is a point on the wall and
${\bf\hat{n}}$ is unit vector normal to the wall.  The force on
a matter particle from the wall is repulsive.
The total potential due to an arbitrary network of planar walls is the sum 
$\Phi({\bf x}) = \sum_i \phi({\bf x};\,{\bf\hat{n}}_i, {\bf x}_i)$. 
Because the potential (\ref{eq:onewall}) grows linearly at
large distances, the walls outside the sphere cannot be neglected in
the calculation of $\Phi$. On the other hand, it is clear that the
contribution of the walls lying outside the event horizon has to
vanish. To model this effect, we introduce a regulator which is
spherically symmetric with respect to the observer and cuts off the effects of
the walls beyond certain radius $\tilde{R} > R$. (Notice that such a
regulator is also necessary in the usual Newtonian treatment of a
matter dominated FRW cosmology.) The dependence on the regulator will
disappear in the final results.

To see how the CMBR anisotropy generated by the walls depends
on their distribution in space, let us consider a ``one-dimensional'' 
toy model in which all the walls are planar and perpendicular to the $x$
axis. The anisotropy, defined in this model as the temperature
difference for the photons arriving from the $\pm x$ directions and measured
by an observer at $x=0$, is
\begin{equation}
  \label{eq:delta}
  \Delta\Phi = \Phi(R) - \Phi(-R).
\end{equation}
Let $N$ denote the number of walls between $x=0$ and $x=R$. Then, we obtain
\begin{equation}
  \label{eq:pot}
  \Phi(R) = F \, R + 4 \pi N G_N \sigma \left( R - \frac{1}{N} 
     \sum_{i=1}^N x_i \right),
\end{equation}
where 
\begin{equation}
  \label{eq:force}
  F = 2 \pi G_N \,\left( - \sum_{i=1}^\infty \sigma f(x_i) + 
      \sum_{i=1}^\infty \sigma f(x_{-i}) \,\right) 
\end{equation}
is the gravitational force exerted by the walls on a unit mass 
placed at the origin, and $x_i$ denote the positions of 
the walls to the right ($i>0$) or to the left ($i<0$) of the observer. The 
regulator function $f$ in Eq.~(\ref{eq:force}) is even, $f(x)=f(-x)$, and
smoothly cuts off the force due to the walls beyond the observer's event 
horizon: $f(x)=1$ for $|x|<\tilde{R}$, $f(x)\rightarrow 0$ for 
$|x|>\tilde{R}$. Using Eq.~(\ref{eq:pot}) and the analogous expression for
$\Phi(-R)$, we can obtain simple analytic estimates of the observed 
anisotropy. In our estimates, we will always assume $N \gg 1$; clearly, 
this condition has to be satisfied for the Universe to be even approximately
homogeneous on large scales. In fact, the CMBR bound derived below will
{\it require} $N$ to be large, so our analysis is self-consistent.

If the walls form a perfectly regular lattice structure with period $L \ll R$,
the leading dependence on $R$ in Eq.~(\ref{eq:delta}) cancels out and one
finds~\cite{ZKO}
\begin{equation}
  \label{eq:ZKO}
  \Delta \Phi \sim 2\pi G_N\sigma L.
\end{equation}
This is exactly what one expects on physical grounds. Indeed, on
scales much larger than $L$, a regular wall structure behaves like
uniform dark energy, with its Newtonian potential approaching a
symmetric parabola; all deviations from the parabola occur because of
the granularity of the structure on scales $\sim L$. However,
Eq.~(\ref{eq:ZKO}) only holds if the wall structure is perfectly
regular on all scales up to the present size of the horizon\footnote{Although
the final result is regulator-independent, the cancellation of the
$R$-dependent terms requires a conspiracy between the ``nearby'' walls
at $x < R$ and the ``distant'' walls at $x \sim \tilde{R}$.}. Both
causality considerations and the inherent randomness of the system
make it extremely hard to believe that such a regular structure can be
realized physically.

In realistic models, the main contribution to the anisotropy comes not 
from the granularity of the wall network, but from long-wavelength
fluctuations of the effective average energy density of the walls due to 
deviations from perfect regularity. As an illustration, consider an example 
in which the walls are displaced from their lattice positions 
by random amounts $|\delta x_i|\sim L/2$. In this case, we find
\begin{equation}
  \label{eq:sqrtN}
  \Delta \Phi \sim 2\pi G_N\sigma (L R)^{1/2},
\end{equation}
a much stronger anisotropy than indicated by Eq.~(\ref{eq:ZKO}). Even this 
model, however, is hardly realistic, since it requires that the number of 
walls to the left and to the right of the observer be identical. Removing 
even a single wall on one side of the observer leads, on average, to an 
anisotropy 
\begin{equation}
  \label{eq:N}
  \Delta \Phi \sim 2\pi G_N\sigma R \sim 2\pi G_N\sigma N L.
\end{equation}
In the absence of long-distance correlations between wall positions, the 
average difference in the number of walls to the left and to the right of the
observer is of order $\sqrt{N}$, leading to 
\begin{equation}
  \label{eq:anisotr_random}
  \Delta \Phi \sim 2\pi G_N\sigma R^{3/2} L^{-1/2} \sim 
   2\pi G_N\sigma N^{3/2} L.
\end{equation}

The preceding discussion can be summarized by writing the anisotropy
in the form
\begin{equation}
  \label{eq:anisotr_general}
  \Delta \Phi  \sim 2\pi G_N\sigma N^{\nu} L,
\end{equation}
where the value of the exponent $\nu$ depends on the details of the wall
configuration. What values of $\nu$ correspond to realistic
domain wall networks? To answer this question, let us consider
a field theory with $N_v$
distinct vacua. Domain walls form during a cosmological phase
transition due to the fact that the field may choose different vacua
in causally disconnected regions of space. Immediately after the phase
transition, each causally connected region contains either no walls,
with probability $p \sim 1/N_v$, or a single wall with probability
$1-p$.  First, consider the case when the wall network does not evolve
(apart from conformal Hubble stretching) after the transition. In this
case, the CMBR anisotropy induced by the walls today can be estimated
as
\begin{equation}
  \label{eq:anisotr_p}
  \Delta \Phi \sim 2\pi G_N\sigma N^{3/2} L \sqrt{p(1-p)}.
\end{equation}
Thus, for moderate $p$, a non-evolving domain wall network induces the
anisotropy of the size indicated by Eq.~(\ref{eq:anisotr_random}).
The same estimate applies if the network does evolve, but the
evolution does not make the structure more regular. On the other hand,
if more regular wall configurations are favored dynamically and emerge
as the network is evolving, the induced anisotropy could be weaker.
While we are not aware of any numerical simulations that conclusively
demonstrate such behavior in a given model, this remains a logical
possibility. A reasonable lower bound on the induced anisotropy is
provided by Eq.~(\ref{eq:N}), since even a single ``defect'' in the
regular wall structure inside the present Hubble volume would induce
an anisotropy of that size.  Below, we will derive the bounds on the
parameters of the model using Eq.~(\ref{eq:anisotr_p}) (the
``non-evolving network'' case, $\nu=3/2$) and Eq.~(\ref{eq:N}) (the
``regular network'' case, $\nu=1$). The constraints in any realistic
model are expected to lie between these two limiting cases.

While we have used a one-dimensional toy model to derive Eqs.~(\ref{eq:ZKO}
--- \ref{eq:anisotr_p}), it is possible to show that the same estimates 
hold for three-dimensional domain wall networks with the corresponding 
regularity properties. Thus, the temperature anisotropy created by domain 
walls inside a sphere of radius $R$ for the non-evolving network case
can be written as 
\begin{equation}
  \label{eq:anisotr_T}
  \frac{\delta T}{\langle T\rangle} = 2\pi a\,G_N\sigma N^{3/2} L 
\sqrt{p(1-p)},
\end{equation}
where $a$ is a numerical coefficient of order unity that depends on the 
detailed properties of the network, $L$ is the average 
separation between the walls, and $N=R/L$ is the average
number of walls crossed by a CMBR photon traveling from the surface of
the sphere to the observer. For a regular network, we obtain
\begin{equation}
  \label{eq:anisotr_N}
  \frac{\delta T}{\langle T\rangle} = 2\pi a\,G_N\sigma N L.
\end{equation}
This anisotropy is generated by the fluctuations of the wall number density 
at large (Hubble) distance scales. Predicting the power spectrum of 
the wall-induced anisotropy, as well as possible deviations from 
gaussianity, would require detailed knowledge of the network geometry. 

To be a viable dark energy candidate, the network of domain walls must
have the average energy density $\rho = 3\sigma/L = \Omega_w \rho_{\rm crit}$,
where $\rho_{\rm crit} = 3H_0^2/8\pi G_N$ is the current critical density,
and $0.6 \lesssim \Omega_w \lesssim 0.8.$ (We assume that the walls are the 
only form of dark energy present: for example, the cosmological constant 
vanishes.) To avoid conflict with the
precise measurements of the CMBR anisotropy~\cite{CMBR}, 
the wall contribution should be at most at the level
$\delta T/T \sim 10^{-6}$. Using Eq.~(\ref{eq:anisotr_general}), we find
\begin{equation}
  \label{eq:boundNgeneral}
  N\gtrsim (2\times 10^5 a b^2)^{1/(2-\nu)},
\end{equation}
where $b$ is the radius of the sphere in Hubble units, $b=RH_0 \approx z_0$.
In the case of a non-evolving network, we obtain 
\begin{equation}
  \label{eq:boundN}
  N\gtrsim 3\times 10^{10} \,a^2 b^4 p(1-p),
\end{equation}
\begin{equation}
  \label{eq:tension}
  \sigma \lesssim \,
{2 \times 10^{-16}  \over a^2 b^3 p (1-p)} \;{\rm GeV}^3,
\end{equation}
\begin{equation}
  \label{eq:separation}
  L \lesssim {0.15 \over a^2 b^3 p (1-p)} \; {\rm pc}, 
\end{equation}
while in the case of a regular network
\begin{equation}
  \label{eq:boundN1}
  N\gtrsim 2\times 10^{5} \,a b^2,
\end{equation}
\begin{equation}
  \label{eq:tension1}
  \sigma \lesssim \,
\frac{1}{a b} \;4 \times 10^{-11}\;{\rm GeV}^3,
\end{equation}
\begin{equation}
  \label{eq:separation1}
  L \lesssim \frac{1}{a b} \; 3\times 10^4\;{\rm pc}.
\end{equation}
Since the wall energy density becomes comparable to that of
matter at $z \simeq 0.5$, we will use the value $b=0.5$ for our
numerical estimates. The other
two parameters entering the above bounds, $a$ and $p$, depend on many
factors, such as the geometry of the wall network, the properties of the
underlying field theory model, etc. We will keep the dependence on these
parameters explicit throughout the discussion. 

Our calculation has only included the anisotropy due to the walls at
low redshifts, where Newtonian approximation is applicable. The effects
of walls at higher redshifts could be included using the full general 
relativistic solution for the metric perturbation created by the walls in 
the FRW Universe~\cite{VS}. In this formalism, the redshift is computed
using the well known Sachs-Wolfe formula~\cite{SW}:
\begin{equation}
  \label{eq:SW}
\left(\frac{\Delta T}{T}\right)_{SW}\,=\,\Phi |_e^o \,-\,
{\bf v}\cdot{\bf e} |_e^o\,-\,\frac{1}{2}\int_e^o h_{\rho\sigma,0}
\dot{x}^{(0)\rho}\dot{x}^{(0)\sigma} d\xi,
\end{equation}
where $h_{\mu\nu}$ is the metric 
perturbation due to the walls\footnote{Explicitly, $ds^2=a^2(\eta)
(g^{(0)}_{\mu\nu}+h_{\mu\nu})dx^\mu dx^\nu$, where $\eta$ is conformal 
time.}, $\Phi=h_{00}/2$ is the ``conformal'' Newtonian potential,
and $x^{(0)}=($const$+\xi, \xi {\bf e})$ is the unperturbed photon path. 
The subscripts $e$ and $o$ refer to the ``emitter'' (i.e., a point on the 
surface of last scattering) and the observer, respectively. Note that in
an expanding Universe, even static domain walls induce a time-dependent
metric perturbation, and the third term in Eq.~(\ref{eq:SW}) does not vanish.
It was shown in~\cite{VS} that in matter-dominated Universe, the anisotropy 
due to a wall at a redshift $z_w$ scales as $(1+z_w)^{-3/2}$ (the redshift 
of a wall is defined as the redshift corresponding to the point on the wall 
closest to the observer.) Moreover, a constant average number density of 
walls in comoving coordinates implies that $dN/dz_w \propto (1+z_w)^{-3/2}$.
Both these effects lead to a severe suppression of the effects of the
walls at high redshifts. We have checked that the calculation of 
Ref.~\cite{VS} leads to the same order-of-magnitude estimates~(\ref{eq:N}) 
and~(\ref{eq:anisotr_T}) for the anisotropy created by the walls as the
Newtonian treatment of our paper, justifying the latter.

\section{Implications} 

Let us discuss the implications of the 
CMBR bounds~(\ref{eq:boundN}---\ref{eq:separation1}) for the field-theoretic 
models responsible for the walls. Domain walls necessarily appear 
in theories with spontaneously broken discrete symmetries. If all the
dimensionless parameters of the model are of order one, the wall tension 
is determined by the symmetry breaking scale $v$. The 
bounds~(\ref{eq:tension}) and~(\ref{eq:tension1}) then imply
\begin{equation}
  \label{eq:scale}
  v \lesssim \, {10 \over a^{2/3} p^{1/3} (1-p)^{1/3}}\; {\rm ~keV}
\end{equation}
and
\begin{equation}
  \label{eq:scale1}
  v \lesssim \, \frac{4\times 10^2}{a^{1/3}}\; {\rm ~keV}
\end{equation}
for the non-evolving and the regular network cases, respectively.
Note that the dependence on $a$ and $p$ is rather mild. In supersymmetric 
models, such low energy scales can be generated naturally and be radiatively 
stable, provided that the SUSY breaking is communicated to 
the field(s) responsible for domain walls only by gravitational interactions.
In this case, the natural value of the discrete symmetry breaking 
scale is given by $v \sim F/M_{Pl}$, where $\sqrt{F}$ is the 
scale at which SUSY is broken. The constraints~(\ref{eq:scale}) 
and~(\ref{eq:scale1}) are satisfied if $\sqrt{F} \lesssim 10^4 - 10^5$ 
TeV, which is allowed phenomenologically
if the breaking is mediated to the visible sector (Standard Model fields and 
their superpartners) by gauge interactions~\cite{GM}. In this respect, the 
domain wall models of dark energy are much more attractive than the
scalar-field quintessence models which contain a superlight scalar
field with a mass of order $10^{-33}$ eV. In the latter case, it is
difficult to understand how such a low energy scale can arise from particle 
physics and not be destabilized by radiative corrections. Moreover, 
the superlight scalar will in general mediate a phenomenologically
problematic new long-range force~\cite{Carroll}. 

The simplest field theory model in which domain walls arise contains  
a single real scalar field with a $Z_n$-invariant potential. (A 
well-known example is an axion of Peccei-Quinn 
models~\cite{Sikivie}.) Numerical simulations~\cite{PRS,PRS2} 
and analytic calculations~\cite{Hindmarsh}
show that the wall network of this model enters the so-called 
{\it scaling regime} shortly after its formation. In this regime, there
is only one wall (or at most a few walls) per horizon volume at any given 
time. This is in gross contradiction with the bound~(\ref{eq:boundN}) and, 
more generally, with the whole concept of the isotropic FRW cosmology. The 
problem could be avoided by introducing a very large number ($\sim 10^8)$
of independent scalar fields, or considering models with a very large 
number of vacua. (In the context of the axion, the latter possibility 
would correspond to a humongous value of the color anomaly of the 
Peccei-Quinn symmetry, $N_{PQ} \sim 10^8$.) Needless to say, both 
possibilities are extremely unattractive.    

The scaling evolution of a domain wall network relies on the fact that
the network can disentangle as fast as allowed by causality. In models
with more complicated vacuum structure, the disentanglement process
could be slowed down, since domains of the same vacuum would generally
be well separated from each other in space. While no convincing
simulations of domain wall networks exhibiting such behavior exist at
present, a similar phenomenon has indeed been observed in the
simulations of cosmic string networks \cite{PenSpergel}.
Alternatively, the walls could experience strong friction forces, for
example due to their interaction with the particles of dark
matter~\cite{Massarotti}. In both cases, one expects substantial
deviations from the scaling law.  The networks of this kind are
referred to as {\it frustrated}.  For frustrated networks, the number
of walls in the present horizon volume depends on their evolution, and
predicting it would require detailed numerical simulations. On the
other hand, one can obtain a simple analytical upper bound on this
number as a function of the wall formation temperature $T_f$. Assuming
that the walls form during the radiation dominated epoch, and that the
fields of the hidden sector responsible for the walls are in thermal
equilibrium at approximately the same temperature as the visible
sector fields\footnote{An upper bound on the temperature of the
  hidden sector fields is provided by Big Bang Nucleosynthesis. Unless
  the hidden sector possesses a large number of degrees of freedom,
  this bound can be satisfied with $T_{\rm hid}$ somewhat lower than,
  but of the same order as, $T_{\rm vis}$.}, we obtain
\begin{equation}
  \label{eq:formT}
  N_h \lesssim z^{-1}(T_f)\,{T_f^2 \over M_{Pl} H_0},
\end{equation}
where $z(T)$ is the redshift corresponding to the given temperature. (This
bound is saturated by a non-evolving network, which in the present context 
can also be termed {\it maximally frustrated}.) In a 
model with no unnatural dimensionless parameters, $T_f \sim v$, and 
Eq.~(\ref{eq:scale}) leads to $N_h \lesssim 10^5 \, a^{-2/3} p^{-1/3}
(1-p)^{-1/3}$ for the non-evolving case. This value does not contradict the 
CMBR constraint~(\ref{eq:boundN}) for reasonable values of $a$ and $p$ (for 
example, $a \sim p \sim 0.1$.) For the case of a regular network, 
Eqs.~(\ref{eq:scale1}) and~(\ref{eq:formT}) imply that $N_h \lesssim
6 \cdot 10^6 \, a^{-1/3}$, although the true bound is probably somewhat 
lower since some of the walls are likely to be destroyed during the 
evolution leading to a regular structure. In any case, the bound seems 
to be compatible with the CMBR constraint~(\ref{eq:boundN1}). Thus, we 
conclude that in the presence of
frustration, it should be possible to build realistic models of domain
wall dark energy without any fine-tuning (apart from the tuning required to 
cancel the cosmological constant.) It would be very interesting to 
find explicit field theory models leading to a frustrated wall
network. 

The upper bounds on $N$ derived in the previous paragraph and the lower 
bounds in Eqs.~(\ref{eq:boundN}) and~(\ref{eq:boundN1}) have to be nearly 
saturated to be
compatible with each other for reasonable values of $a$ and $p$.
This observation leads to two interesting predictions. First, the wall network
has to be close to maximal frustration, or, equivalently, be nearly static. 
Second, the CMBR anisotropy induced by the walls should be close to the
current bounds, and therefore improved measurements of the anisotropy 
have a good chance of detecting the wall contribution if this model is 
realized.   

\section{Equation of state}

It is well known that the equation of state of a maximally frustrated
(static) network of planar domain walls is $w=-2/3$. Above, we have argued
that a domain wall network has to be close to the maximally frustrated  
regime to play the role of dark energy and be consistent with the CMBR
anisotropy limits. One may worry, however, that even small deviations from
this regime may cause a large change in the value of $w$, making it an
unacceptable dark energy candidate. This turns out not to be the case.
Let us demonstrate that the wall equation of state lies in 
the allowed range $-2/3\le w\le -1/2$, regardless of the evolution.

Energy conservation for domain walls reads 
$d(\rho V)=-p dV-\delta E$, where $\delta E$ is the energy lost 
by the network in the process of evolution. (The energy can be 
radiated away in form of gravitons, elementary excitations of the fields
responsible for the walls, etc.) Let us tentatively
set $\delta E=0$. The wall equation of state can then be found from
the dependence of the energy density of the wall network $\rho_w$ on
the scale factor of the Universe $a$,
\begin{equation}
  \label{eq:wfromrho}
  w = -1-{1\over3}{d \ln \rho_w \over d \ln a}.
\end{equation}
The number
of walls in a fixed comoving volume scales as $\eta^{-\alpha}$, where
$\eta$ is conformal time, $d\eta=dt/a(t)$. When the wall network is
static, $\alpha=0$. On the other hand, by causality there must be at
least one wall per horizon volume at any given time, implying $\alpha
\le 1$. The physical density of the wall rest energy therefore scales
as $\propto a^{-1}\eta^{-\alpha}$. Neglecting the kinetic energy
of the walls, we obtain
\begin{equation}
  \label{eq:walls2}
  w = -\frac{2}{3}+{\alpha\over3}\,{d \ln \eta \over d \ln a}.
\end{equation}
During the matter dominated epoch the total pressure of the Universe
$p_{\rm tot}=0$, while after the wall energy takes over $p_{\rm tot}$
becomes negative. It can then be shown that in an expanding Universe
$0<d \ln \eta / d \ln a\leq 1/2$, and therefore 
\begin{equation}
  \label{eq:wrange}
  -\frac{2}{3}\leq w \leq -\frac{1}{2}.
\end{equation}
The lower limit is achieved when the domain wall network is static,
while the upper limit corresponds to the scaling regime.  Notice that
even in the scaling regime, while the Universe is matter dominated,
the wall energy density varies as $\propto a^{-3/2}$.  Hence,
regardless of the details of the evolution, the wall network
eventually comes to dominate the Universe and its equation of state
always lies within the range allowed by the analysis of~\cite{concord}.
Of course, as already mentioned, the network must be nearly maximally 
frustrated to satisfy the isotropy constraint, so we generally expect 
$w \simeq -2/3.$ In this case, the more restrictive bound obtained 
in~\cite{PTW} is also satisfied\footnote{Recent studies disfavor $w=-2/3$ 
at 95\% confidence level~\cite{Add}}. 

In the above derivation, we have neglected the energy loss by the wall 
network $\delta E$ and the kinetic energy of the walls $E_{\rm kin}$.
Obviously, both approximations are valid for static
networks. Moreover, numerical simulations~\cite{PRS,PRS2} 
show that even in the scaling regime $\delta E \simeq 0$ and $E_{\rm kin} 
\ll E_{\rm rest}$ throughout the evolution. This justifies our assumptions. 
(Notice also that the condition $\delta E=0$ can be relaxed; the effect of 
finite $\delta E/\delta a$ is to shift $w$ closer to $-2/3$, and 
Eq.~(\ref{eq:wrange}) still holds.) On the other hand,
the condition $p_{\rm tot}\leq 0$ is essential for
deriving Eq.~(\ref{eq:wrange}). For example, in the radiation
dominated Universe the wall equation of state is $w=-1/3$ in the
scaling regime. 

\section{Conclusions} 

We have considered the 
constraints on domain wall models of dark energy from the observed
near-isotropy of the CMBR. We have shown that these constraints can be 
satisfied by a strongly frustrated domain wall network. The scale of 
spontaneous symmetry breaking responsible for the walls is expected to 
lie in the 10-100 keV range, and can arise naturally in supersymmetric 
theories. This makes these models quite attractive from the particle 
physics point of view.

Domain wall models of dark energy have important observational
predictions. The dark energy equation of state is predicted to be
close to $-2/3$. This value can be clearly distinguished from the case
of the cosmological constant by the SNAP experiment~\cite{SNAP}. The
CMBR anisotropy induced by the walls is likely to be close to the
current bounds, and could be observable in the near future. In
general, inhomogeneities of the wall distribution are also expected to
induce peculiar velocity flows on large scales. However, this effect is
small: a network satisfying the 
CMBR constraints will produce velocities of order 300 m/s, which is
not in conflict with current observations. 

The most important outstanding issue in making domain wall dark energy 
models fully realistic is finding explicit field theories which lead to 
highly frustrated wall networks. Frustration could arise as a result of 
the complex dynamics of the system. This possibility
can only be addressed by numerical simulations. Such simulations are 
also necessary to make more detailed predictions of the phenomenological 
signatures of the walls, such as the power spectrum of the wall-induced CMBR 
anisotropies. At this time, we are aware of only one numerical study of 
frustrated networks~\cite{Kubotani}, whose usefulness is severely limited by 
its insufficient dynamical range. Clearly, further work in this 
direction is necessary. An interesting alternative possibility is 
to consider domain wall networks whose evolution is slowed down by their 
interaction with dark matter. This idea was introduced in~\cite{Massarotti}.
In the specific model of Ref.~\cite{Massarotti}, the average wall velocity 
is determined by the ratio of wall and dark matter energy densities. In our 
case, this ratio is of order one, and the mechanism does not work. However, 
this class of ideas certainly deserves further investigation.

\section*{Acknowledgments}

We thank G.~Gabadadze, J.~Moore, D.~N.~Spergel and M.~White for helpful 
discussions. A.F. is supported by the Keck Foundation; H.M. is supported by 
the National Science Foundation under grant PHY-0098840;  
M.P. is supported by the Director, Office of Science, Office of High
Energy and Nuclear Physics, of the U. S. Department of Energy under 
Contract DE-AC03-76SF00098. 

\appendix

\setcounter{equation}{0}
\section{Newtonian treatment of the photon redshift}
\label{sect:appendix}

The rigorous way of computing the temperature anisotropy acquired by
the photons on the way from the surface of last scattering to the observer
is by using the techiques of general relativity.  On the other hand,
the anisotropy induced when the photons are travelling inside a sphere of 
a radius corresponding to $z_0\lesssim 1$ and centered on the observer can 
be simply estimated using Newtonian theory. (The estimate is expected to
be accurate up to corrections suppressed by powers of $z_0$.) 
To illustrate how the Newtonian theory can be applied to give the
correct redshifts, accurate to the second order in $v/c$, let us
explicitly compare the redshifts of a photon emitted at some distance
$R$ computed by the two methods in the simplest case of exactly 
homogeneous, uniformly expanding Universe.  

From the point of view of general relativity, the ratio of the emitted
and observed frequencies, $\omega_1/\omega_2$ is simply the
relative change of the scale factor of the Universe $a$ during the
time between emission and absorption,
\begin{equation}
\omega_1/\omega_2=a_2/a_1.
\end{equation}
If the Universe is filled with a component at critial density with the
equation of state $w$, the scale factor has a power law dependence on
time, $a\sim t^n$, where $n=2/[3(1+w)]$.
The velocity with which the emitter is receding from the observer at
the time of emission $t_1$ is
\begin{equation}
v=R\frac{\dot{a}(t_1)}{a(t_1)}=R\frac{n}{t_1}.
\end{equation}
The time it takes for the light to travel to the observer, 
$\Delta t\equiv t_2-t_1$,  can be found from the equation
\begin{equation}
\int_{t_1}^{t_2}\frac{dt}{a(t)}=\frac{R}{a(t_1)}.
\end{equation}
Upon integration, we find 
\begin{equation}
R=(t_2^{1-n}t_1^n-t_1)\frac{1}{1-n},
\end{equation}
so that 
\begin{equation}
t_2=\left(\frac{(1-n)R+t_1}{t_1^n}
\right)^{1/(1-n)}.
\end{equation}
Using this result, we find for the ratio of the scale factors
\begin{equation}
\frac{a (t_2)}{a(t_1)}=\left(\frac{t_2}{t_1}\right)^n=
\left((1-n)\frac{R}{t_1}+1\right)^{n/(1-n)},
\end{equation}
or, in terms of the recess velocity $v=n R/t_1$,
\begin{equation}
\frac{a (t_2)}{a(t_1)}=
\left(1+v\frac{(1-n)}{n}\right)^{n/(1-n)}.
\end{equation}
Finally, the ratio of the emitted to absorbed frequency is
\begin{eqnarray}
\label{eq:redshiftGR}
\frac{\omega (t_2)}{\omega(t_1)}&=&
\left(1-v\frac{(n-1)}{n}\right)^{n/(n-1)}
\simeq 1-v+\frac{1}{2 n}v^2+...\nonumber\\
&\simeq& 1-v+\frac{3(1+w)}{4}v^2+...
\end{eqnarray}

Now let us compute the frequency redshift using Newtonian
gravity. For that we take the observer to be at ``the center of the
Universe'', with the surrounding matter inducing a gravitational field
\begin{equation}
{\bf g}(r)=-G\frac{4\pi}{3}(\rho+3 p)r{\bf\hat{r}}.
\end{equation}
The frequency shift has two components, the kinematic Doppler shift
and the blueshift because the photon falls into the potential well.
The Doppler shift is simply given by
\begin{equation}
\label{eq:DopplerNewt}
\omega_2^{\rm Doppler}=\frac{\omega_1(1-v)}{\sqrt{1-v^2}}
\simeq\omega_1 (1-v+v^2/2+...)
\end{equation}
The gravitational blueshift is computed as follows
\begin{eqnarray}
\label{eq:gravNewt}
\frac{(\omega_2-\omega_1)^{\rm Grav}}{\omega_1}&=&\Delta\phi=
-\int_0^R g(r) dr\nonumber\\
&=&G\frac{4\pi}{3}(\rho+3 p)\frac{R^2}{2}.
\end{eqnarray}
Since in the flat Universe the recession velocity obeys $v(R)^2/2=G
M(R)/R=4\pi G \rho R^2/3$,  Eq.~(\ref{eq:gravNewt}) could be rewritten
as 
\begin{eqnarray}
\label{eq:gravNewt2}
\frac{(\omega_2-\omega_1)^{\rm Grav}}{\omega_1}
=(1+3 w)\frac{v^2}{4}.
\end{eqnarray}

We observe that, by combining Eqs.~(\ref{eq:DopplerNewt}) and
(\ref{eq:gravNewt2}), we recover precisely the result of
Eq.~(\ref{eq:redshiftGR}). Thus, in a homogeneous Universe, the
Newtonian calculation of the photon redshifts due to the expansion of
the Universe is correct up to and including the second order terms
in $v$. In Section II, we use the Newtonian picture to estimate the 
anisotropies due to the presense of domain walls. Since the walls only
become dominant for $z\sim 1$, this approximation should be sufficiently 
good for an order-of-magnitude estimate. This intuition is confirmed by
studying the full general relativistic calculation of Ref.~\cite{VS}.

\end{document}